\documentclass[fleqn,twoside]{article}
\usepackage{espcrc2}

\usepackage{graphicx}
\usepackage[figuresright]{rotating}

\def\simge{
    \mathrel{\rlap{\raise 0.511ex
        \hbox{$>$}}{\lower 0.511ex \hbox{$\sim$}}}}
\def\simle{
    \mathrel{\rlap{\raise 0.511ex 
        \hbox{$<$}}{\lower 0.511ex \hbox{$\sim$}}}}

\newcommand{\AmS}{{\protect\the\textfont2

  A\kern-.1667em\lower.5ex\hbox{M}\kern-.125emS}}

\title{
\vskip -110pt
\vskip 45pt
Lattice calculation of the lowest-order hadronic contribution to
         the muon anomalous magnetic moment }

\author{
T. Blum
\address{Physics Department, University of Connecticut, Storrs, 
               CT 06269-3046}
\address{RIKEN BNL Research Center, Brookhaven National
               Laboratory, Upton, NY 11973 USA}
}

\begin{document}

\begin{abstract}
I present quenched domain wall fermion and 2+1 flavor improved Kogut-Susskind fermion calculations of the hadronic
vacuum polarization which are used to calculate the ${\cal O}(\alpha^2)$ hadronic contribution to the anomalous magnetic moment of the muon. Together with previous quenched calcuations,
the new results confirm that in the quenched theory the hadronic
contribution is signifcantly smaller ($\sim 30\%$) than the value obtained from the total cross section of $e^+e^-$ annhilation to hadrons.
The 2+1 flavor results show an increasing contribution to $g-2$ as
the quark mass is reduced.
\end{abstract}

\maketitle

\section{Introduction}
The anomalous magnetic moment of the muon (g-2) has been measured to incredible accuracy at Brookhaven National Lab's E861 experiment
\cite{Bennett:2004pv} and has also been calculated precisely in the
Standard Model using the dispersion relation for the hadronic vacuum polarization, $\Pi(q^2)$, and the total cross section for $e^+e^-$ annihilation to hadrons (isospin symmetry allows the decay rate for $\tau$ leptons to be used as well)\cite{Davier:2003pw,Ghozzi:2003yn}. 
The theory calculations agree with experiment at roughly 2.7 and 1.4 standard deviations, depending on whether solely $e^+e^-$ data are
used or if $\tau$ decay data are included as well. 

A while ago it was shown how to compute the lowest order (in $\alpha$) hadronic contribution  to $g-2$
completely from first principles, using lattice gauge theory to calculate the hadronic vacuum polarization in conjunction with continuum perturbation theory in the QED coupling constant\cite{Blum:2002ii}. In this work we continue along this path, past new high statistics quenched domain wall fermion calculations, to 2+1 flavor improved Kogut-Susskind fermion calculations on large volumes $V \simle (3.5$ fm)$^3$ and with small quark masses ($m_l\equiv (m_u+m_d)/2 \approx m_s/10$).
In the latter case I use configurations from the MILC collaboration.
The 2+1 flavor calculation is not complete, so I do not give numbers for g-2 in this case. After presenting results, I discuss the work that still needs doing and prospects for a precision lattice calculation of
g-2.
 
The vacuum polarization is defined from the two-point correlation function of the electro-magnetic current 
\begin{eqnarray}
\int_{x} d^4x\,J^\mu(x) J^\nu(y) e^{iq\cdot(x-y)} &=&\\\nonumber
(q^2\delta^{\mu\nu}-q^\mu q^\nu)\Pi(q^2),&&
\end{eqnarray}
where $q$ is photon momentum. See \cite{Blum:2002ii,Gockeler:2003cw} for details of the lattice calculation.
The central idea is to calculate $\Pi(q^2)$ on the lattice, and fit it
to obtain a continuous function that can be readily integrated (numerically) with a know function from continuum perturbation theory
from $q^2=0$ up to a cut-off $\sim (1/a)^2$, the rest of the integral being done using perturbation theory. For the quenched case, the
(continuum) form of $\Pi(q^2)$ is known and provides an ansatz for
the fit\cite{Gockeler:2003cw}.
\begin{eqnarray}
\Pi(q^2) &=& \frac{f_V^2}{q^2+m_V^2} + C \ln{(a^2(q^2+\mu^2))},
\end{eqnarray} 
where the first term comes from the vector meson bound state (delta 
function) and the second from a two particle continuum (cut, $q^2 \ge 
\mu^2$). Note that $f_V$ and $m_V$ can be extracted in the usual way from the zero momentum correlator as was done in \cite{Gockeler:2003cw}
which leads to a more accurate fit, or treated as free parameters (here, I treat $f_V$ as a free parameter and take $m_V$ from\cite{Aoki:2002vt}). This ansatz fits the data well and leads to (statistically) 
accurate results that are extrapolated to $q^2=0$, which is important 
because the low $q^2$ region dominates the one-loop integral that 
gives the hadronic contribution. For the 2+1 flavor case there is, of course, 
no such phyical ansatz, excluding the experimental one which is, after
 all, the thing we're trying to compute in the first place. In this case I try
 several forms: the pole fit just described, a simple polynomial, a log 
 for the low $q^2$ region, and combinations of these.

\section{Results and Discussion}

\begin{figure}[h]
\vspace*{-0.cm}
\includegraphics[width=2.95in]{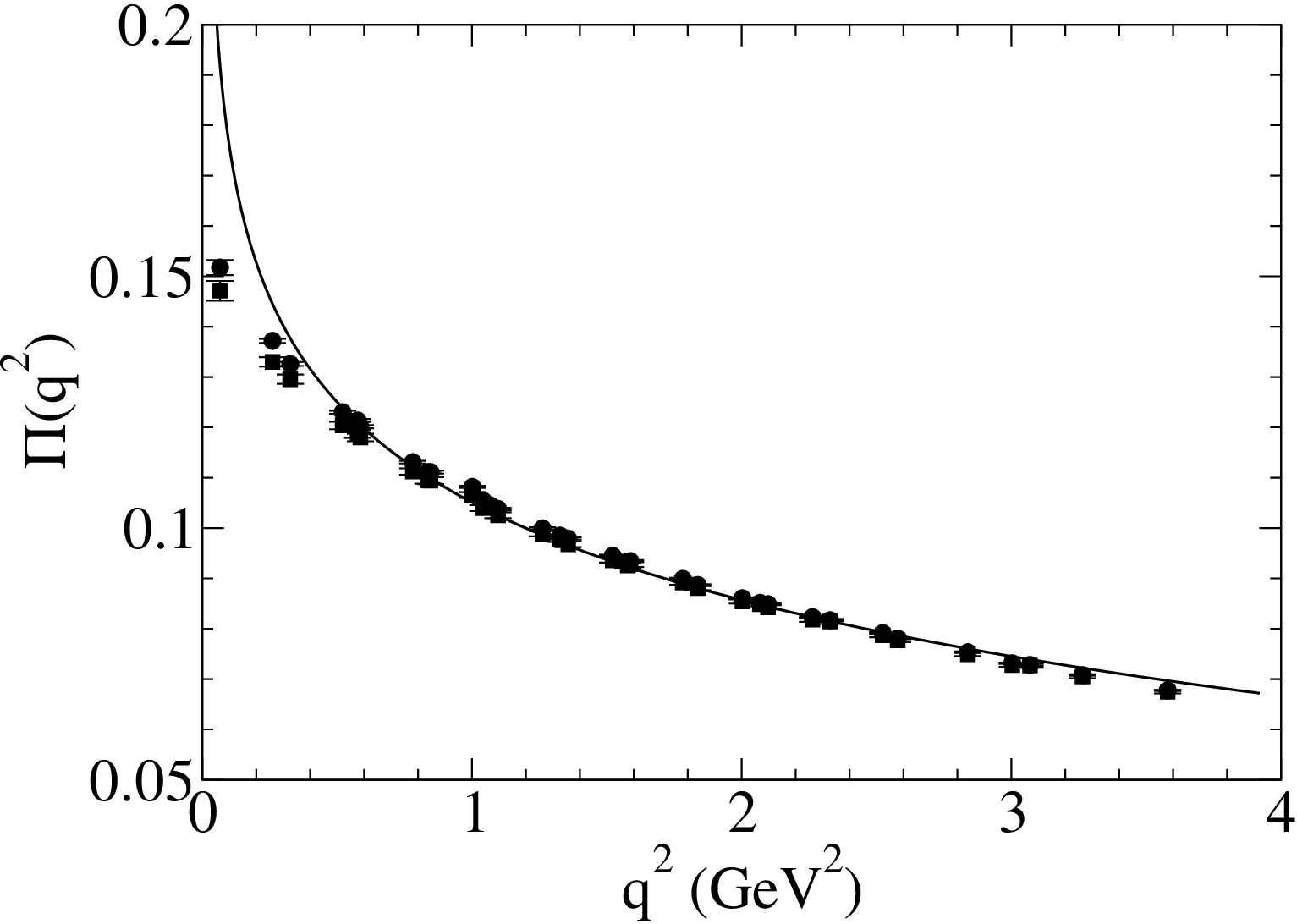}
\vspace*{-.50cm}
\caption{The hadronic vacuum polarization for quenched domain wall fermions, $a^{-1}=1.3$ GeV, $m_{val}=0.02$ (circles) and 0.04 (squares). The line shows the result from continuum 3-loop perturbation theory\cite{Chetyrkin:1995ii} for comparison.}
\label{fig:dwf}
\end{figure}

The quenched calculations with domain wall fermions were done with inverse spacing $a^{-1}\approx 1.3$ and 2 GeV on $16^3\times 32$ lattices, and $m_{val}=0.02$ and 0.04. $\Pi(q^2)$ is shown in Figure \ref{fig:dwf},
and values of $a_\mu=(g-2)/2$ for a single quark with unit charge are sumarized in Table 1. After including the u, d, and s quark charges, the total contribution for three degenerate quarks is consistent with previous quenched results\cite{Blum:2002ii,Gockeler:2003cw}, confirming that the hadronic contribution to $g-2$ is significantly less ($\sim 30\%$) than in the real world\cite{Davier:2003pw}. Thus it is probably not worthwhile to pursue further quenched calculations designed to eliminate systematic errors due to  finite volume, non-zero lattice spacing, and unphysically large quark mass (which has been done to some extent in 
\cite{Blum:2002ii,Gockeler:2003cw}).

Instead, I have started a calculation on the 2+1 flavor lattices 
from the MILC collaboration that were generated using $a^2-$tad fermions (here I omit the Naik term for the valence quarks\cite{Blum:2003se}). Initial results were presented last year\cite{Blum:2003se}, and currently $\Pi(q^2)$ is being calculated on new $40^3\times96$
lattices with $m_l=0.0031$. Of course, these are very aggressive parameters, so the lattice generation is somewhat slow. At the time of the meeting some 84 configurations existed on which $\Pi(q^2)$ was calculated from a point-split current from a single site (and its nearest neighbors). These were separated by six trajectories and so are probably not independent. To improve statistics, I have begun calculating on a time-slice that is one-half the lattice size distant from the first. In addition
the MILC collaboration plans to at least triple the length of the 
evolution ($\sim 3000$ trajectories). 

In Figure 2 $\Pi(q^2)$ is shown for  
$m_{val}=m_{l}=0.0031$ ($40^3\times96$) and 0.0062 
($28^3\times96$), or $0.1$ and 0.2 times the strange quark mass, 
respectively. Note that the large volumes and time sizes used here lead to very small values of $q^2$ which is quite important. 
The data show a slight increase as $q^2\to 0$ as $m_l$ 
decreases by a factor of two. While it appears small, a slight increase
in this region changes the contribution to $g-2$ significantly, so the fit in this region must be quite accurate and precise. In Figure 3 I show covariant polynomial fits to $\Pi(q^2)$ which tend to under-predict the data as $q^2\to 0$ (the $\chi^2$ of the cubic and quartic fits is acceptable, though). Pole fits like Eq. 2 do a bit worse, and a log fit does about the same. It seems that a better fit ansatz is needed.

Chiral perturbation theory offers the means to understand the mass dependence and the small $q^2$ behavior of $\Pi(q^2)$. This result does not appear in the literature, to the best of my knowledge, and
even if it did, to be effective in this case, it would probably have to be augmented with so-called staggered chiral perturbation theory (S$\chi$PT) \cite{Bernard:2001yj}
to obtain accurate results. Indeed, recently the MILC collaboration has performed a fit to all of their 2+1 flavor data for the pseudo-scalar decay constant using S$\chi$PT\cite{Aubin:2004fs}. 
The combination of a large data set
and S$\chi$PT allowed for a $\sim$ 1-2\%
determination of $f_K/f_\pi$ which is quite remarkable. 
Since the vacuum polarization can be measured
accurately on these same lattices, perhaps a similar fit will 
prove to be as precise here as there. I am now considering such an analysis for $\Pi(q^2)$.

\begin{table}[h]
\caption{The magnetic anomaly $a_\mu$ for a single flavor with unit charge from quenched domain wall fermions.}
\begin{tabular}{ccccc}
\hline
$a^{-1}$ & $m_{val}$ & configs & $(\alpha^2)$ $a_\mu^{\rm had}$ \cr
\hline
1.31 & 0.02 & 337 & 750(35) x10$^{-10}$&\cr
1.31 & 0.04 & 337 &669(23) x10$^{-10}$&\cr
1.98 & 0.02 & 299 &730(51) x10$^{-10}$&\cr
\hline
\end{tabular}
\vskip -.25in
\end{table}

\begin{figure}[h]
\vspace*{-0.cm}
\includegraphics[width=2.95in]{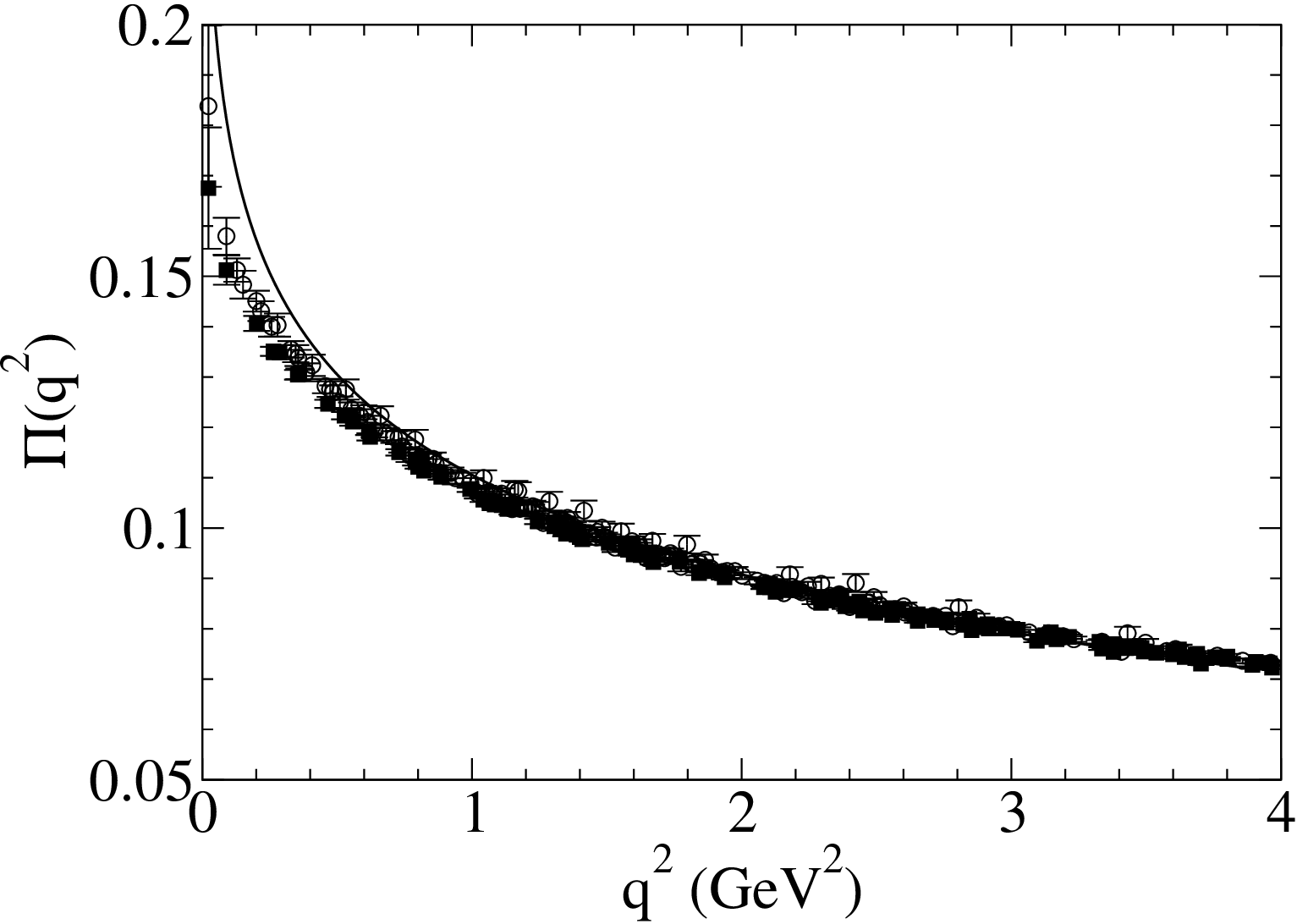}
\vspace*{-.50cm}
\caption{Same as Figure 1 except for 2+1 flavor improved Kogut-Susskind fermions, $a^{-1}\approx 2.3$ GeV, $m_s=0.031$, $m_{val}=m_l = 0.0031$ (open circles) and 0.0062 (filled squares).}
\label{fig:ks}
\vskip -.5in
\end{figure}

\begin{figure}[h]
\vspace*{0.2cm}
\includegraphics[width=2.95in]{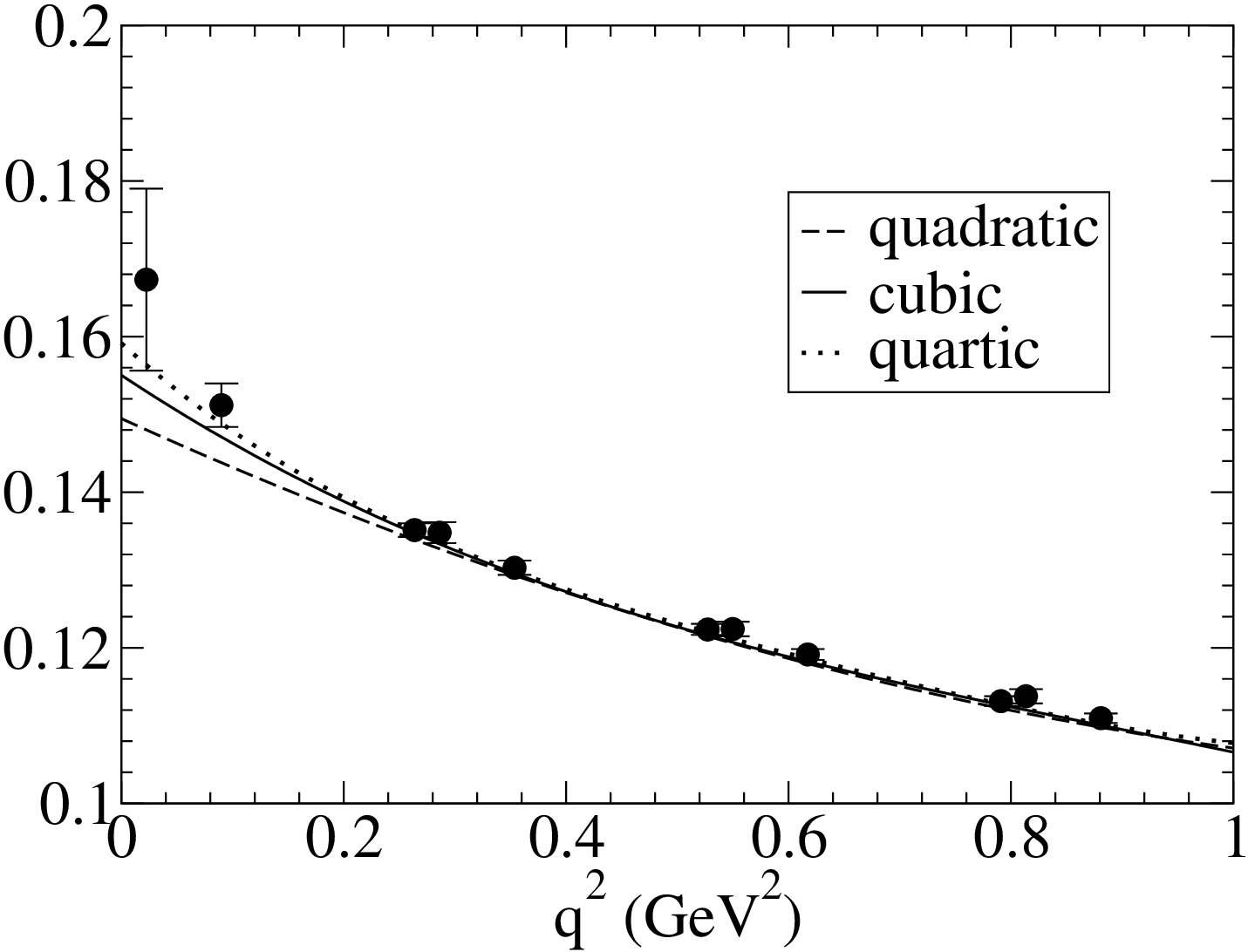}
\vspace*{-.50cm}
\caption{Polynomial fits to $\Pi(q^2)$ for the 2+1 flavor case with $m_{val}=0.0062$.}
\label{fig:ks fit}
\vskip -.25in
\end{figure}

\section*{Acknowledgment}
\noindent
This work was partly supported by the LDRD program of Brookhaven
National Laboratory, Project No. 04-041. The domain wall fermion calculations were performed on the QCDSP supercomputer at the
RIKEN BNL Research Center.
The staggered fermion calculations were performed on Seaborg (IBM-SP2) at NERSC using gauge configurations from the MILC 
collaboration and code based on 
the MILC code (version 6) for which I am grateful.
I thank RIKEN and the US DOE for providing the resources to 
accomplish this work.

\end{document}